\documentclass[journal]{IEEEtran}

\usepackage[ruled, linesnumbered]{algorithm2e}
\usepackage{subcaption}

\usepackage{amssymb}
\usepackage{amsmath, amsthm, amsfonts}
\usepackage{amsfonts} 
\usepackage{soul}
\usepackage{color}

\usepackage{graphicx, wrapfig, subcaption, setspace, booktabs}

\usepackage{multicol}
\usepackage{afterpage}
\usepackage{tabularx}
\usepackage{tikz}

\usepackage{cite}
\usepackage{mdwmath}
\usepackage{textcomp}
\usepackage{soul}
\usepackage{nomencl}
\usepackage{array}  
\usepackage{setspace}  
\usepackage{xcolor} 
\usepackage{bm}  

\ifCLASSINFOpdf
\else
\fi

\def\OJlogo{\vspace{-4pt}\includegraphics[height=18pt]{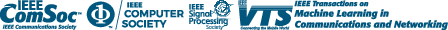}}
\def\seclogo{\vspace{10pt}\includegraphics[height=18pt]{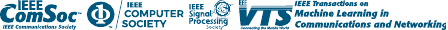}}

\def\authorrefmark#1{\ensuremath{^{\textbf{#1}}}}

\begin{document}

\title{Anticipating Optical Availability in Hybrid RF/FSO Links Using RF Beacons and Deep Learning}

\author{Mostafa Ibrahim\authorrefmark{1,2}, 
Arsalan Ahmad\authorrefmark{2},~\IEEEmembership{Senior~Member,~IEEE},
Sabit~Ekin\authorrefmark{3},~\IEEEmembership{Senior~Member,~IEEE},
Peter LoPresti\authorrefmark{4},~\IEEEmembership{Senior~Member,~IEEE},
Serhat Altunc\authorrefmark{5}, 
Obadiah Kegege\authorrefmark{5}, 
and John~F.~O'Hara\authorrefmark{2},~\IEEEmembership{Senior~Member,~IEEE}

\thanks{\authorrefmark{1}Department of Engineering Technology, Texas A\&M University, Texas, USA. ~\authorrefmark{2}School of Electrical and Computer Engineering, Oklahoma State University, Oklahoma, USA.~~\authorrefmark{3}Department of Electrical and Computer Engineering, Oklahoma State University, Oklahoma, USA.~~\authorrefmark{4}Department of Electrical and Computer Engineering, University of Tulsa, Tulsa, OK, USA.~~\authorrefmark{5}Goddard Space Flight Center NASA, Greenbelt, Maryland, USA.}
\thanks{Corresponding author: Mostafa Ibrahim (email: mostafa.ibrahim@tamu.edu).}
\thanks{This work was supported in part by the National Aeronautics and Space Administration under Grant 80NSSC20M0214; in part by the U.S. Department of Energy, Office of Science, Office of Advanced Scientific Computing Research, under Award DE-SC0023957.}
\thanks{This work has been submitted to the IEEE TMLCN for possible publication.  Copyright may be transferred without notice, after which this version may no longer be accessible.}
}


\maketitle

\begin{abstract}
Radio frequency (RF) communications offer reliable but low data rates and energy-inefficient satellite links, while free-space optical (FSO) promises high bandwidth but struggles with disturbances imposed by atmospheric effects. A hybrid RF/FSO architecture aims to achieve optimal reliability along with high data rates for space communications. Accurate prediction of dynamic ground-to-satellite FSO link availability is critical for routing decisions in low-earth orbit constellations. In this paper, we propose a system leveraging ubiquitous RF links to proactively forecast FSO link degradation prior to signal drops below threshold levels. This enables pre-calculation of rerouting to maximally maintain high data rate FSO links throughout the duration of weather effects. We implement a supervised learning model to anticipate FSO attenuation based on the analysis of RF patterns. Through the simulation of a dense lower earth orbit (LEO) satellite constellation, we demonstrate the efficacy of our approach in a simulated satellite network, highlighting the balance between predictive accuracy and prediction duration. An emulated cloud attenuation model is proposed which provides insight into the temporal profiles of RF signals and their correlation to FSO channel dynamics. Our investigation sheds light on the trade-offs between prediction horizon and accuracy arising from RF beacon proximity, achieving a prediction accuracy of 86\% with 16 RF beacons.

\end{abstract}

\begin{IEEEkeywords}
Satellite Communication, Hybrid RF/FSO, Ray Tracing, Low-Earth Orbit (LEO), FSO availability, and cloud attenuation. 
\end{IEEEkeywords}


\section{Introduction}

Global IP traffic growth has seen an enormous increase over the past few years courtesy of applications like big data, E-gaming, audio \& video streaming, and cloud computing \cite{telefonica2023}. Optical communication has served as a pivotal technology in efficiently managing the escalating traffic demands across various network segments, including core and access, offering unparalleled terabit per second capacity. However, the utilization of optical fiber communication in geographically challenging regions with sparse populations is hindered by substantial upfront deployment capital expenditure (CAPEX) and other legal \& logistical challenges. Satellite communication has emerged as a viable solution in situations where providing traditional terrestrial communication infrastructure is challenging, offering an alternate means of bridging the communication gap \cite{telecomreview2023}. Radio frequency (RF) and free-space optical (FSO), as well as hybrid RF/FSO architectures have emerged as part of this vision.

There are also numerous upcoming space and lunar missions that underscore the need for advancements in hybrid RF/FSO satellite communication links. These missions, ranging from lunar exploration to establishing sustainable outposts, highlight the growing demand for reliable and high-capacity communication systems. 
Notable future missions include NASA’s Artemis program \cite{angelopoulos2014artemis}, which aims to return humans to the Moon and establish a sustainable presence by the end of the decade. NASA had been using standard high data rate RF links for Earth, Space, and Explorations missions and leading the way for current and future Optical communications missions.
Optical communications can be viewed as ambitious but certainly realistic considering the multiple optical comm missions; Lunar Laser Communications Demonstration (LLCD) \cite{NASA2015LLCDFactSheet}, NASA Laser Communications Relay Demonstration (LCRD) \cite{NASA_LCRDOverview}, NASA's TeraByte InfraRed Delivery (TBIRD) \cite{NASA2022FastestLaserLink}, Orion Artemis II Optical Communications System (O2O) \cite{NASA_O2O2024}, Integrated LCRD LEO User Modem and Amplifier Terminal (ILLUMA-T) \cite{NASA_ILLUMAT2023}, and Deep Space Optical Communications (DSOC) \cite{NASA_DSOC2023}. Additionally, other nations have an interest in the Lunar regime and considering RF/FSO systems. European Space Agency’s Lunar Pathfinder mission \cite{giordano2022lunar} intends to provide communication services to lunar missions. China’s lunar exploration plans \cite{li2019china} involve both robotic and crewed missions, with the goal of building a lunar research station. India’s Chandrayaan-3 mission \cite{durga2023contextual}, aiming for lunar landing and exploration, and Russia’s Luna-Glob \cite{smirnov2013luna}, an orbiter and lander mission, are also in the pipeline. The complexity and distance of these missions necessitate robust communication links, making the development of efficient hybrid RF/FSO systems a critical component of success.


Radio-frequency (RF) communication has been the main workhorse for mission communications in almost every satellite deployed to date due to its reliability over a variety of conditions. They are proven, robust, and reliable, but also slow and energy-inefficient compared to the FSO links. Optical communication (also known as laser communication or free-space optical (FSO) communication) can provide high-data-rate links; however, restrictions imposed by Earth weather and pointing errors limit its exclusive use \cite{8799061}. Therefore, a hybrid RF/FSO approach is the most practical option to provide critical high-data-rate, reliable, and efficient communication links and network architecture for space communications. Commensurately, there has been growing interest in hybrid RF/FSO satellite communication systems to harness the high bandwidth capacity of optical links while overcoming their vulnerability to disruption \cite{Raza}. 

Numerous tutorials and surveys in recent years have covered non-terrestrial 5G/6G and advanced cellular networks \cite{lin20215g}, \cite{rinaldi2020non}, \cite{cheng20226g}, \cite{wang2019convergence}, point-to-point communication \cite{saeed2021point}, and artificial intelligence in satellite communication. 
The application of artificial intelligence (AI) in non-terrestrial networks (NTN), space-ground links, and satellite-air-terrestrial network technology trends are extensively reviewed in \cite{zhang2020survey}. The work in \cite{naous2023reinforcement} provides a survey on leveraging machine learning to improve NTN network key performance indicators, backhaul reliability, data integrity, and spectrum management. Additionally, \cite{fourati2021artificial} offers a comprehensive review of AI in satellite communication, focusing on optimizing beam hopping, network traffic forecasting, channel modeling, scintillation detection, and interference and energy management.
As another example of machine learning applications in satellite communications, specifically between high-altitude platforms (HAP) and LEO constellations, the work in \cite{lee2020integrating} explores deep reinforcement learning for analyzing atmospheric conditions.
In \cite{arani2021fairness}, the authors address the challenge of providing backhaul connectivity to UAVs via LEO platforms, employing Q-Learning to solve a multi-armed bandit problem focused on user fairness and minimizing terrestrial base station inefficiency.

Numerous studies have explored integrating RF and FSO for robust satellite connectivity \cite{Abadi,Halim}. The work in \cite{kaushal2016optical} presents a thorough survey of the challenges faced in FSO communication systems for both ground-to-satellite/satellite-to-ground and inter-satellite links. While hybrid RF/FSO satellite networks have advantages, further innovations in FSO link availability, seamless switching between RF and FSO links, and channel modeling are needed to fully realize the potential gains.

The principal challenge RF/FSO hybrids are meant to address is that FSO link availability suffers in the presence of fog, clouds, rain, and atmospheric turbulence.  In a hybrid RF/FSO scheme with a large LEO constellation, numerous satellites can serve to re-route traffic around weather-related obstructions and close the link with the nearest available ground station (destination), thereby maintaining the high-data rate of purely FSO communications.  A subset of satellites can also be equipped with long range FSO terminals to simultaneously service distant (e.g. Lunar) missions.  Alternatively, the hybrid LEO nodes can switch between FSO and the more robust RF transmission when the FSO link signal-to-noise (SNR) vacillates around a critical threshold.  This binary decision scheme is regularly found in the extant literature but poses a problem in that switching may occur too late to maintain continuous connectivity since satellites require significant time in positioning, acquisition, and tracking (PAT) tasks.  With the fast-changing nature of ground-to-space links, it becomes increasingly important to anticipate and pre-calculate the best possible re-routing or switching schemes to maximize link uptime and data rates.   

In this paper, we propose utilizing an RF/FSO hybrid architecture in a unique way that aims to increase overall FSO link availability by using RF links as atmospheric sensors that can proactively predict FSO link availability before sudden interruptions occur.  The key contributions are:

\begin{itemize}

    \item Proposing a hybrid RF/FSO architecture for satellite communications that leverages RF links to predict FSO availability. This system significantly enhances the reliability of satellite communications, a critical factor in guaranteeing uninterrupted communication amidst dynamic weather disruptions.
    \item  Devising an architecture consisting of a main ground station that communicates with a constellation of LEO satellites via an FSO link, supported by a network of RF links between ground beacons and the satellites. The technique uses signals from RF beacons to predict the degradation of FSO links up to a minute before it occurs.
    \item  Designing a supervised machine learning model that analyzes RF signal patterns to predict FSO link attenuation. This model translates the complex relationship between RF signal strength and atmospheric conditions into a predictive tool for FSO link reliability.
    \item Demonstrating and validating the proposed techniques through simulations of the LEO constellation network under an emulated cloud attenuation model. The modeling provides vital insights into expected system performance under different conditions, such as the dynamic changes in cloud cover that can obstruct FSO links.
    
\end{itemize}
The remainder of this paper is organized as follows. Section II discusses the system model, including details on RF and FSO attenuation and prediction of temporal calculations. Section III describes the implementation of the machine learning model. Section IV presents the simulation scenario, cloud emulation algorithm, and results. Finally, Section V concludes the paper.

\begin{figure}
    \centering
    \includegraphics[width=.35\textwidth]{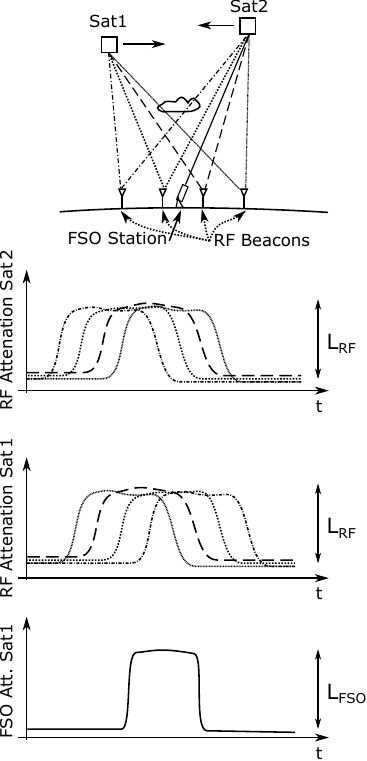}
    \caption{System Overview: Illustrates Sat1 and Sat2 orbiting in opposite directions, with a cloud blob attenuating RF beacon and FSO station signals, alongside corresponding attenuation curves.}
    \label{fig:symmodel}
\end{figure}

\section{System model}

The system is composed of an FSO link between a main ground station and a constellation of LEO satellites. This is assisted by a network of RF links between ground beacons and the LEO satellite constellation, as shown in Fig \ref{fig:symmodel}. The figure illustrates the FSO link as a solid line and the RF links as dotted or dashed lines of different styles, both on the geometric and the graph illustrations. 

Fig. \ref{fig:symmodel} shows that the positioning and direction of different satellites relative to clouds and RF ground beacons can create a pattern in the RF attenuation $L_{RF}$ that is related and can predict the FSO attenuation, $L_{FSO}$, ahead of time. This is due to the fact that they are both correlated through the position of the cloud, its dimensions, and water content. The values of the FSO and RF attenuations $L_{FSO}$ and $L_{RF}$ are dictated by the corresponding cloud attenuation models.

\begin{figure*}[h]
    \centering
    \includegraphics[width=.58\textwidth]{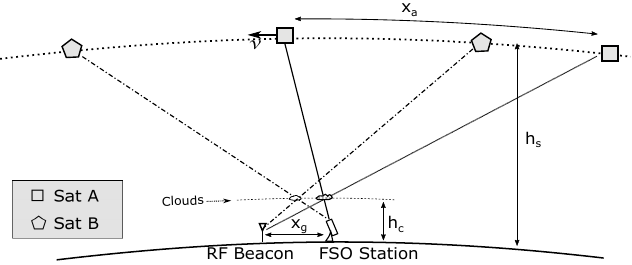}
    \caption{Geometry of early attenuation prediction: The figure depicts $Sat A$ at two time instances, illustrating cloud interference with RF beacon signal and subsequent FSO signal disruption after traversing distance $x_a$. $Sat B$ has two different instances, hence marking a different point on the cloud layer.}
    \label{fig:tauest}
\end{figure*}

\subsection{RF and FSO Attenuation:}
\label{rffsoA}
RF models describing cloud attenuation use cloud coverage parameters such as water content and cloud thickness to create reasonably accurate predictions of the received power levels \cite{alawadi2012investigation}.  Dissanayake-Allnutt-Haidara (DAH) is a rain-attenuation propagation model \cite{dissanayake1997prediction} valid up to 35 GHz. The Salonen-Uppala cloud attenuation model \cite{salonen1991new} has been used by Cost 255 for frequencies up to 160 GHz. Based on that model, the ITU recommendation \cite{ITU-R-P.840}, valid up to 200 GHz, uses a Rayleigh scattering approximation and expresses the attenuation in terms of total water content per unit volume.
The cloud attenuation according to the ITU model is:
\begin{equation}
    A_c= \dfrac{L~ k_l}{\sin(\Theta) }~~~(dB),
\end{equation}
where $\Theta$ is the elevation angle, $L$ in the total column content of liquid, reported in the recommendation charts, and the specific attenuation coefficient $k_l$~((dB/km)/(g/m$^3$))  is:
\begin{equation}
    k_l=\dfrac{0.819 f }{\epsilon'' (1+\eta^2)},
\end{equation}
where $\epsilon = \epsilon'+j\epsilon''$ is the complex dielectric permittivity of water, $f$ is the frequency in (GHz), and $\eta=\frac{2+\epsilon'}{\epsilon''}$.

For the FSO link, fog and cloud-induced attenuation can be estimated using a visibility range parameter $V$ (km) \cite{awan2009cloud}. Kim's model \cite{kim2001comparison} can be used to define the attenuation coefficient $\omega$~(dB/km) as in \cite{nadeem2009weather}:
\begin{equation}
    \omega = \dfrac{3.91}{V} \left( \dfrac{\lambda_{FSO}}{550} \right)^{-x}, 
\end{equation}
where $\lambda_{FSO}$ is the wavelength of the optical link, and $x$ is the particle size coefficient which can be determined from experimental data in \cite{middleton1952vision}.
\begin{equation}
x = 
\begin{cases} 
1.6 & ; V > 50 \text{ km } (\text{high visibility}) \\
1.3 &; 6 \text{ km} < V \leq 50 \text{ km } (\text{average visibility}) \\
0.585V^{\frac{1}{3}} & ; V < 6 \text{ km } (\text{low visibility})
\end{cases}    
\end{equation}


The FSO attenuation will be several times larger than the RF attenuation, necessitating the operation of the RF link in a higher frequency band (above 100 GHz) to increase sensitivity to cloud presence. The RF beacon signal is a single tone for ease of power level capturing. Hence, a longer duration of RF signal capture can be used to enhance the accuracy of the estimated attenuation level.
Our study's link budget and attenuation level calculations rely on a ray tracing propagation model. Differing from the aforementioned statistical models, this method allows for precise modeling of wireless signal propagation in intricate environments, providing real-time insights into signal strength and attenuation. We have integrated a cloud model into our ray tracing simulation to assess its impact on link attenuation. The model executes ray tracing calculations to ascertain the link attenuation along the direct path between the satellite and the RF ground beacons.

The ray tracing geometric model employs a large-scale fading approach, which effectively captures overall propagation characteristics over extended distances. This consideration allows us to discern spatial variability in signal strength within the cloud. To simplify the channel model and due to the lower RF attenuation compared to the FSO link, we have opted to exclude Fresnel zone calculations. Consequently, we do not account for diffraction losses from the edges of the clouds. This decision is warranted based on our assessment of the negligible impact of Fresnel zones on overall attenuation for our specific scenario.

\subsection{Prediction Temporal Calculations}
In this section, we examine the geometric aspects of the problem and explore the temporal relationship between RF and FSO attenuation processes. To provide a first-order approximation, we utilize the two-dimensional projection shown in Fig. \ref{fig:tauest} without considering the Earth's curvature.

The proportions in the figure are schematic rather than scaled. The LEO satellite altitudes, denoted as $h_s$, typically range from 250 km to 2000 km, whereas the altitude of the cloud layer, represented as $h_c$, ranges from 2 km to 10 km.  We assume a separation $x_g$ between the RF and FSO stations, usually spanning hundreds of meters. The points at which we assess attenuation are positioned at the cloud layer's altitude.

To ensure reliable communication, we limit the analysis to satellite links at a specific minimum elevation angle above the horizon. This minimum elevation angle boundary creates a volume shaped as an inverted cone with the vertex at the FSO station, and the base facing the open space. This volume is the operation region for our study. 
The satellite covers a distance $x_a$ in time $\tau$, with velocity $v$.
The velocity can be determined using the equation:

\begin{equation}
    v=\sqrt{\frac{ G M}{r}},
\end{equation}
where $G$ represents the gravitational constant, $M$ is the mass of the Earth, and $r$ is the radius of the orbit.

In this context, we define the lookahead duration, denoted as $\tau$, for a specific point in the sky. This duration represents the temporal difference between two distinct positions of a satellite: the position where the satellite intersects the line connecting the RF beacon and obstructing cloud and the subsequent position where the satellite intersects the line connecting the FSO station and the same cloud. Essentially, $\tau$ measures the time interval required for the satellite to move from the intersection point of the RF path to the intersection point of the FSO path at that specific location in the sky. It can be determined as follows:
\begin{equation}
\tau=\dfrac{(h_s-h_c)~x_g}{v~h_c }
\label{eq:tau}
\end{equation}


In our two-dimensional diagram, we simplify by depicting the two satellite instances and the two ground stations within the same plane that includes the zenith point. However, this is a simplification for illustrative purposes; actual satellite orbits are three-dimensional, with each satellite having its unique maximum elevation angle with respect to the ground station. Satellites with a maximum elevation angle of less than \(90^\circ\) do not transit through the zenith of the FSO station. Despite this, the relationship presented for \(\tau\) remains valid even if the plane of operation in the figure is tilted with angle $\theta_{inclination}$, because the factor $ \left( 1/\cos(\theta_{inclination}) \right) $ will appear in both \(h_s\) and \(h_c\), thus canceling out in the Eq. \ref{eq:tau}. Although the satellite's speed vector may vary due to the changing elevation angle, this effect is relatively minor and can be neglected for the purposes of our model.

We can realize that different points in the cloud layer correspond to different satellite positions relative to a specific RF beacon and satellite trajectory. Thus a comprehensive model is required to address these interrelated parameters effectively and forecast FSO link quality accurately. Moreover,  given the diverse directional movement of satellites in their orbital spheres, RF beacons need to encircle the FSO station, as we shall show in the coming sections.

\begin{figure}[h]
    \centering  \includegraphics[width=.25\textwidth]{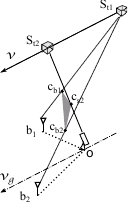}
    \caption{ The geometric relationship for early attenuation prediction in 3D space. The figure shows the satellite's movement vector relative to beacons $b_1$ and $b_2$, and the resulting $\triangle c_{s2} c_{b1} c_{b2}$ within the plane containing the cloud altitude, where attenuation prediction is approximated.}
    \label{fig:tauest3d}
\end{figure}

Next, we present the geometry of early attenuation prediction in three dimensions (3D), as depicted in Fig. \ref{fig:tauest3d}. In this scenario, the satellite's movement vector $v$ is not aligned with the direction of either beacon $b_1$ or $b_2$; however, it runs parallel to the ground vector $v_g$, which lies between the line segments $\overline{o~b_1}$ and $\overline{o~b_2}$. Consequently, it is impossible to form a plane using the points $\{ b_1, S_{t1}, S_{t2}, O\} $ or the points  $\{ b_2, S_{t1}, S_{t2}, O\} $, as long as time instances are not the same $(t1 \neq t2)$. This geometry implies that the line segments $\overline{b_1~S_{t1}}$ and $\overline{O~S_{t2}}$ do not intersect, and the same is true for $\overline{b_2~S_{t1}}$ and $\overline{O~S_{t2}}$. 
The closest approximation to an intersection point is the triangle formed within the cloud layer, denoted as $\triangle c_{s2} c_{b1} c_{b2}$. In essence, the prediction point in the sky expands into a region rather than a single point. The shape of this triangle is influenced by geometric parameters such as the height of the cloud layer, the vector $\overrightarrow v$, and the lengths of the segments  $\overline{O~b_1}$ and $\overline{O~b_2}$
Despite this, points within this region exhibit a high correlation due to the cloud's shape characteristics. The attenuation at point $c_{s2}$ is partially correlated with the attenuation at  $c_{b1}$, and similarly with point $c_{b2}$. Therefore, our goal is to capture these complex relationships through training our learning model.

Cloud heights, satellite elevation angles, velocity components, and $x_g$ spacings correspond to different $\tau$ values. As a result, the model must be trained with a range of $\tau$ values to account for all possible satellite positions and directions. In the following section, we propose a machine learning model for prediction and design an algorithm that considers the aspects above.
\section{Learning Algorithm Implementation}

\subsection{The Machine Learning Model}
In this section, we propose the utilization of supervised learning to predict the FSO attenuation.
Supervised learning involves a dataset comprising input-output pairs to train a model to associate inputs with corresponding outputs. Here, we employ a neural network as the model, mapping the relationship between the prior RF signal power received from the beacons and the anticipated quality of the FSO link $\tau$ seconds in the future. The learning scheme exploits the correlation-based learning concept.

Correlation-based learning refers to the idea of using correlations between samples or experiences in the learning process. 
By leveraging the neural network's ability to learn and generalize from data, correlation-based learning can provide valuable insights and predictive capabilities in scenarios where traditional analytical models may fall short.

The correlation between the RF signal power and the FSO link availability arises from the geometric configuration of the system and the presence of obstacles such as clouds. By considering factors such as the shape of the cloud, its direction, and speed, as well as the satellite's direction and speed, it is possible to derive an analytical formula for predicting FSO link availability. However, due to the complexity of the problem, we opt to employ a neural network to learn this formula instead.

 The inputs to the neural network consist of the RF power measurements from the beacons and the position of the satellite at different time instances $\tau_1, \tau_2, \ldots, \tau_m$ preceding the prediction of FSO link power level. 
We represent these inputs as measured attenuation levels along with satellite positions and trajectory: $\mathbf{X} = [ LRF_{tx1, \tau_1},~ LRF_{tx2, \tau_1},~ LRF_{txn, \tau_1}, \cdots , ~LRF_{txn, \tau_m},$
$ {R,~ AZ,~EL},~ {\delta AZ,~ \delta EL} ]$. Where (${R, ~AZ,~EL}$) is the position vector ( distance, azimuth, and elevation angles), and the angle increments between two time instances (${\delta AZ,~ \delta EL}$) reflect the direction information. 

 These metrics constitute the features of the neural network. The output of the network is the predicted attenuation of the FSO link. Depending on a predefined attenuation threshold, this output indicates whether it is likely to be operational or not. 
 

\subsection{Experience buffer}
The environment we aim to predict exhibits variability, with the characteristics and direction of clouds evolving over time. This necessitates a dynamic model capable of adapting to such variations. To achieve this, we employ experience buffers to store information from a past fixed duration, from which a batch of samples is randomly selected for model training. Though experience buffers are traditionally linked with reinforcement learning scenarios, their application proves advantageous in our context.
\begin{figure}[h]
    \centering
    \includegraphics[width=.43\textwidth]{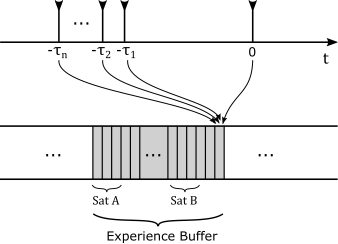}
    \caption{Experience buffer structure.}
    \label{fig:replaybuffer}
\end{figure}
 Fig. \ref{fig:replaybuffer}  illustrates the structure of the experience buffer. Each experience sample contains the FSO measurement at the current time ($t$=0) and the RF measurements captured $\tau$ seconds ago. The measurements are collected and stored at regular intervals, creating a continuous stream of experience samples. The buffer window duration is derived from experiences across several satellites that crossed above a specific elevation angle during that period. These measurements are collected at the satellites and fed back to a central entity, the FSO station. This central entity assumes the responsibility of training the model and either uploading it to the satellites or making predictions on behalf of the satellites.

For the training process, we adopt a mini-batch approach. The model's prediction is represented by the function $F(\mathbf{X}; \theta)$, where $\theta$ denotes the model's parameters. We employ the following loss function for backpropagation:

\begin{equation}
Loss(\theta_t, \mathbf{X}, L_{FSO})= (L_{FSO}-F(\mathbf{X}; \theta))^2,
\label{eq:loss}
\end{equation}
where $N_B$ represents the batch size. The mini-batch update equation in stochastic gradient descent (SGD) is given by:

\begin{equation}
\theta_{t+1} = \theta_t - \alpha \frac{1}{|B|} \sum_{(x, y) \in B} \nabla Loss(\theta_t, \mathbf{X}, L_{FSO}),
\end{equation}
where $\alpha$ denotes the learning rate, and $B$ contains the training samples in the minibatch. It is important to note that we are not restricting the training to SGD, as various optimization algorithms are available for training neural networks, but exploring them falls outside the scope of our work.

In the subsequent section, we conduct simulations using different setups of a LEO satellite constellation, clouds, and beacons to gain deeper insights into our proposed scheme.
\section{Simulations \& results}
We simulate a LEO satellite constellation of 1000 satellites with altitudes ranging from 400 km to 2000 km. A single cloud layer is assumed at an altitude of 8 km. The prediction accuracy is evaluated across different ground beacon distributions.  
From the geometry of the problem, we can understand that a circular positioning of beacons will correspond to a single $\tau$ value for a specific altitude to train for. But because we have a range of altitudes, the trained neural network has to take inputs from several $\tau$ values, as mentioned above.

The ground RF beacons are distributed uniformly in a circle around the FSO station to accommodate the satellites approaching from different orbits. 
The radii of the circles in the simulated scenarios are 250 m, 500 m, 750 m, and 1 km. In the following subsections, we present the cloud emulation process. Then, we demonstrate the signal shapes in a simplified four-beacon model. Finally, we present the prediction accuracy results for the different beacon distributions. 
\begin{figure}[b]
    \centering
    \includegraphics[width=.4\textwidth]{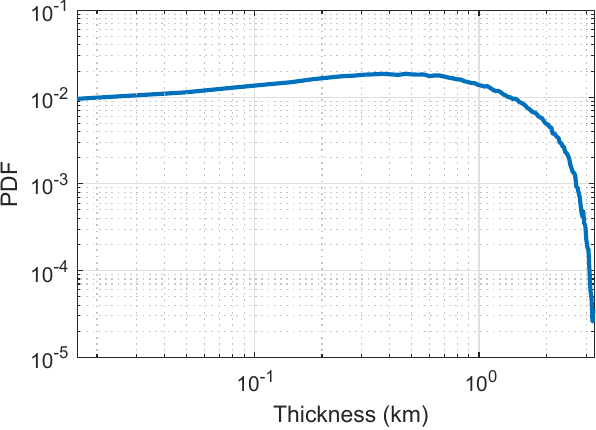}
    \caption{Simulated cloud thickness probability density function.}
    \label{fig:thPSD}
\end{figure}

\begin{figure}[h]
    \centering
    \includegraphics[width=.35\textwidth]{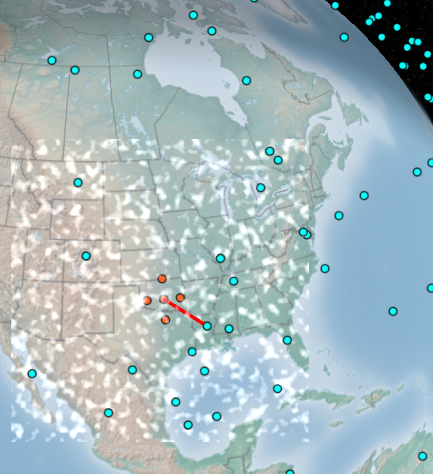}
    \caption{A four-beacon simulation scenario showing the overlayed cloud process (the positioning of the beacons is exaggerated).  Blue dots indicate LEO satellites and red dots indicate ground receivers.  The red line shown connects a single satellite to the FSO ground station.}
    \label{fig:sc1}
\end{figure}

\subsection{Cloud Model}

Our simulations emulate a moving cloud layer with specific directional speed and gradual shape evolution over time. To model graphical clouds, we employ filtered Gaussian noise. Although literature from various disciplines often uses Perlin noise for cloud emulation \cite{5694143,9170750}, we opt for the simplified filtered Gaussian noise model due to its ease of implementation and close results.
\begin{figure*}[t]
    \centering
\includegraphics[width=.99\textwidth]{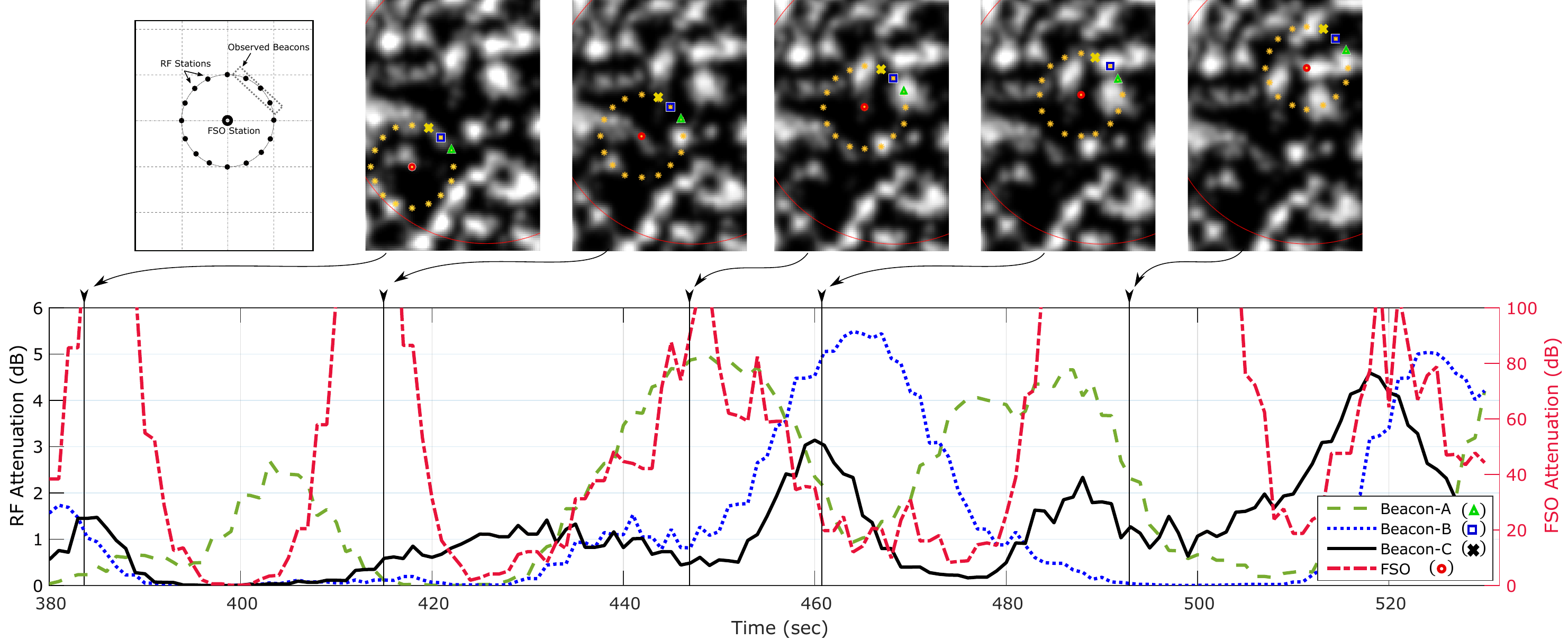}
    \caption{RF and FSO Signal Attenuation Correlation: Displays RF beacons and FSO station signal attenuation due to cloud cover, influenced by satellite movement. The top plot shows station positions; sequential snapshots track cloud impact on signals. The lower plot correlates attenuation patterns with cloud dynamics and satellite path.}
    \label{fig:s390t}
\end{figure*}
The random process formulation is shown in Algorithm 1, while the descriptions of the algorithm notations are mentioned in Table \ref{tab:algonotations}.

\begin{algorithm}
\caption{Cloud Emulation Algorithm}
\label{alg:cloud_emulation}
\SetKwInOut{Input}{Input}
\SetKwInOut{Output}{Output}
\Input{Grid sizes $N_x$, $N_y$; downsample coefficient $u$; speed $(v_x, v_y)$; threshold}
\Output{BlurredNoise for ray tracing channel calculations}

\SetAlgoNlRelativeSize{-1}
\SetAlgoNlRelativeSize{0}

\textbf{Generate} two 2D grids of sizes $[N_x, N_y]$ and $[\frac{N_x}{u}, \frac{N_y}{u}]$ \;
\textbf{Create noise} by combining two Gaussian processes: $\text{noise} = \text{randn}([N_x, N_y]) + \text{interp}(\text{randn}([\frac{N_x}{u}, \frac{N_y}{u}]), N_x, N_y)$ \;
\textbf{Set} $t = 0$ (Initialize the time step) \;
\While{simulation time}{
    $t = t + 1$ (Update the time step) \;
    \textbf{Wrapping} noise grid based on the speed $(v_x, v_y)$: \\
    \For{$i = 1$ to $N_x$}{
        \For{$j = 1$ to $N_y$}{
            $\text{WN}(i,j) = \text{noise}[(i + v_x) \mod N_x~, (j + v_y) \mod N_y]$
        }
    }
    \textbf{Update the noise}: \\
    $\text{noise} = \text{WN}(1-p) + P \text{randn}([N_x, N_y]) + P\text{interp}(\text{randn}([\frac{N_x}{u}, \frac{N_y}{u}]), N_x, N_y)$ \;
    \textbf{Apply thresholding} to create a binary cloud mask: \\
    $\text{Mask} = \text{noise} > \text{threshold}$ \;
    \textbf{Filtering with a blurring kernel for smoothness:} \\
    \For{$i = 1$ to $N_x$}{
        \For{$j = 1$ to $N_y$}{
            $\text{BN}(i, j) = \sum_{m=-m_k}^{m_k} \sum_{n=-n_k}^{n_k} K(m, n) \cdot \text{noise}(i-m, j-n)$
        }
    }
    \textbf{Normalize} cloud thickness average \;
    \textbf{Utilize} the cloud mask for ray tracing channel calculations \;
}
\end{algorithm}

\begin{table}[h]
\centering
\caption{Cloud Emulation Algorithm Notations. }
\label{tab:notations}
\begin{tabular}{ll}
    \textbf{Notation} & \textbf{Description} \\
    \hline
    $\text{randn}([\cdot, \cdot])$ & Generates a matrix of size $[N_x, N_y]$ with \\ & Gaussian random values. \\
    $\text{interp}(\cdot, a, b)$ & Upsamples the matrix $A$ to size $[a, b]$ \\ &  using interpolation. \\
    $K(\cdot, \cdot)$ & Represents the values of the blurring kernel \\ & used in the filtering operation. \\
    $u$ & Downsample coefficient. \\
    $t$ & Time step in the simulation. \\
    $p$ & A parameter to control the amount \\ &  of noise update. \\
    $v_x, v_y$ & Speed components along the $x$ and $y$ \\ & directions, respectively. \\
    $\text{WN}$ & Noise grid after 2D wrapping based on the \\ & speed. \\
    $\text{Mask}$ & Binary cloud mask obtained from  \\ &  thresholding the noise. \\
    $\text{BN}$ & Noise grid after filtering with a \\ & blurring kernel. \\
    \hline
\end{tabular}
\label{tab:algonotations}
\end{table}

It is worth mentioning that the downsampled process creates cloud blobs, while the upsampled process introduces minor variations. The grid wrapping operation is pixel-based and carried out according to the assigned wind speed and direction. Although the thresholding and smoothing steps are additional refinement steps not used in the recursive process, they enhance the simulation's accuracy.

We normalize the values to match cloud thickness statistics in meteorological studies \cite{guillaume2018horizontal}. Fig. \ref{fig:thPSD} depicts the cloud's thickness probability density function (PDF). The cloud layer is superimposed in the simulation environment and rendered for each time frame (Fig. \ref{fig:sc1}). We consider the link's direct paths between the corresponding satellite and the ground station to calculate attenuation values. Based on the RF and FSO attenuation models mentioned in Section \ref{rffsoA}, we assume a 3 dB/km RF attenuation and 300 dB/km FSO attenuation, with the FSO detector able to detect up to 100 dB of attenuation at most.


\subsection{Preliminary Pre-training Results}
In this section, we demonstrate the temporal relationship between RF beacon signal attenuation and FSO signal attenuation. In a setup of 16 RF beacons surrounding the FSO station, we choose the three beacons that are in the direction of the movement of the satellite. The satellite is moving from the southwest toward the northeast. Hence, we are choosing beacons A, B, and C, {as illustrated} in Fig. \ref{fig:s390t}. The first plot on the upper left shows the positions of the stations on the ground, with the RF beacons encircling the FSO station with a radius of 1 Km. The contour of $30^o$ elevation angle from the FSO station, on the cloud layer, is projected on the ground. This is to show that for the following sequence of instances, the ground frame is not moving. 

The sequence of five cloud layer snapshots that follow shows the progression of the cloud cover, depicted in white, against a black background representing a clear sky. The marks \textcolor{orange}{\bm{$\ast$}}, \textcolor{black}{\bm{$\times$}}, \textcolor{green}{\bm{$\Delta$}}, and \textcolor{blue}{\bm{$\square$}} indicate the intersection between the path from the ground beacons and the satellite with the cloud layer. The mark \textcolor{red}{\bm{$\circ$}} represents the intersection point for the FSO station. The snapshots reflect the satellite's movement, hence the movement of the marks. We chose the instances of the snapshots so that the RF and FSO marks cross the same cloud blob, separated by the prediction duration (32 sec).

In the lower plot of Fig. \ref{fig:s390t}, we present the signal attenuation values over time for the three RF beacons A, B, and C, and the FSO station. Intersection points are marked  \textcolor{green}{\bm{$\Delta$}}, \textcolor{blue}{\bm{$\square$}},  \textcolor{orange}{\bm{$\times$}}, and \textcolor{red}{\bm{$\circ$}} consecutively. At instance 1, the beacon marks are crossing a cloud blob that is later crossed by the FSO mark at instance 2. We can also observe on the attenuation curve that the shape and width of the attenuation curves are related to the cloud blob's shape.

We also observe that because the satellite's direction is towards the \textcolor{orange}{\bm{$\times$}} mark (Beacon-C), the attenuation curve of that beacon is the one most closely resembling the shape of the FSO attenuation curve at instance 2. The same pattern is observed in the following sequences: The RF marks at instance 2 cross the blob crossed by the FSO at instance 3. The RF marks at instance 4 cross the blob crossed by the FSO at instance 5. As a result, the duration between time instances 1 and 2 is the same as the duration between 2 and 3, and the same as the duration between 4 and 5. In other words, these durations were chosen to be equal to demonstrate the aforementioned temporal correlation.

We conclude from this section that depending on the satellite's direction, which in our case is in the direction of \textcolor{orange}{\bm{$\times$}}, there is a correlation between the RF attenuation curves and the FSO attenuation curve. The correlation does not have to be a one-to-one match due to the nature of the cloud and its speed. This is especially true if the wind direction differs from the satellite's direction or if the wind speed is high compared to the speed at which the satellite's beacon marks cross the clouds. Additionally, clouds do not only translate but also change shape while moving, which adds extra randomness to the problem. All these parameters will affect the prediction accuracy of our proposed system.




\begin{figure}[h]
\centering
\includegraphics[width=0.3\textwidth]{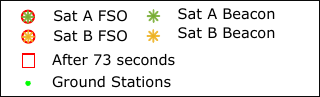}
\begin{subfigure}[b]{0.273\textwidth}
    \centering
    \includegraphics[width=\textwidth]{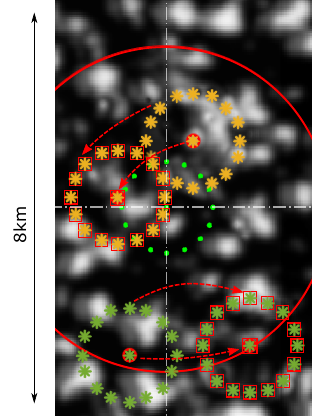}
    \caption{}
    \label{fig:S1000}
\end{subfigure}~~
\begin{subfigure}[b]{0.2\textwidth}
    \centering
    \includegraphics[width=\textwidth]{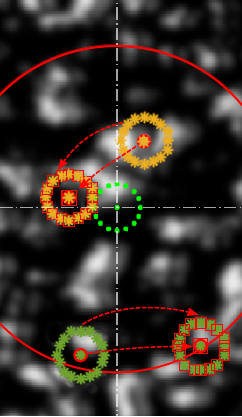}
    \caption{}
    \label{fig:S500}
\end{subfigure}
\caption{Progression of cloud layer intercept contours for the scenarios: (a) 1000 m and (b) 500 m beacon radius. The maps' x and y axes are to scale.}
\label{fig:Overlaps}
\end{figure}

\subsubsection{Geometric demonstration of time progression}
This subsection shows how different scenarios result in different prediction durations. In Fig. \ref{fig:Overlaps}, we show two scenarios, each with the same set of satellites overlayed in two different time frames. The number of beacons is 16, and the $\ast$ marks represent the intersection between an RF link and the cloud layer. The $\circledast{}$ mark is the intersection between the FSO link path and the cloud layer. The RF beacons are positioned circularly with the FSO station at the center. The circles' radii are 500 m in Fig. \ref{fig:S500} and 1 km in Fig. \ref{fig:S1000}.

In Fig. \ref{fig:Overlaps}, the frames are overlayed with transparency of each cloud instance of 40 \%  and 60 \%  for the old and new frames consecutively. Therefore, in the span of 73 seconds, we can observe the shift of the cloud blobs. The 73-second duration is chosen arbitrarily, where at this part of the simulation, only two satellites appeared on the horizon, which is convenient for our demonstration. Same as before, the red circle is the contour of $30^o$ elevation angle above the horizon on the cloud layer from the position of the FSO station. The ground station positions are marked for each scenario, with the FSO station at the center, and the RF beacons surrounding it.

Satellites $Sat A$ and $Sat B$ have different altitudes, therefore they have different speeds and cross different distances for the same time duration. The plots have the same time span between the overlapping frames. We differentiate between the satellites using the yellow and green colors, and we use the square mark to indicate future frames. The cloud layer is shown in the background with black-and-white. Black represents a clear LOS path, and white represents the cloud blockage. 

We can observe in Fig. \ref{fig:Overlaps} that the RF beacon position contours from different frames are separated temporarily from the FSO prediction point with different duration values. These values depend on the radius of the beacon's circle and the satellite altitude, hence speed. In ascending order, the prediction $\tau$ values can be listed as:
\begin{itemize}
    \item $[$Sat A, rad = 500 m, $\tau = 14~ sec]$, 
    \item then $[$Sat B, rad = 500 m , $\tau = 18~ sec]$, 
    \item then $[$Sat A, rad = 1000 m, $\tau = 28~ sec]$, 
    \item then $[$Sat B, rad = 1000 m,  $\tau = 35 ~sec]$.
\end{itemize}
Due to these different satellite speeds, we can have different values of $\tau$ for the same beacon setup. This will require a training network with features that cover the $\tau$ range of all satellites for a specific setup.
In the next section, we show each scenario's choice of the $\tau$ value ranges and how this reflects on the training input features.


\begin{figure*}[t]
    \centering
\includegraphics[width=.99\textwidth]{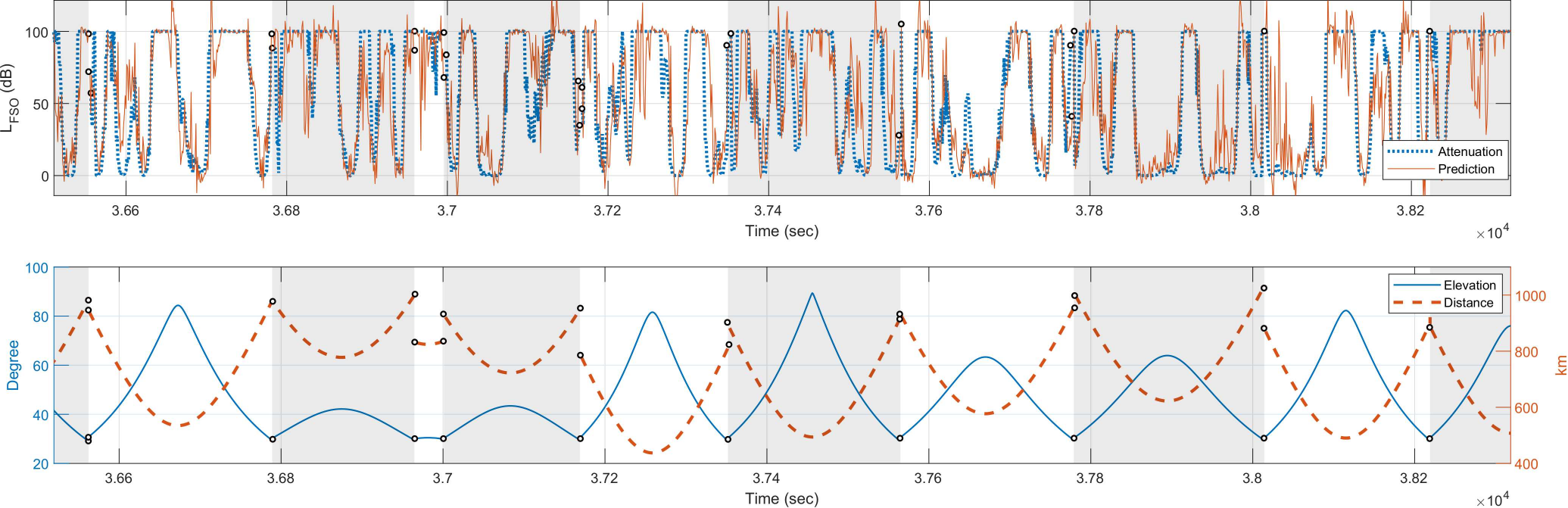}
    \caption{ FSO attenuation and prediction for the [250 m, 16 beacon] scenario. The results are from several satellites, with the corresponding distance and elevation angle plotted on the same time axis.  }
    \label{fig:pred250}
\end{figure*}

\subsection{Evaluation of Training Parameters for Different Scenarios}

In this section, we evaluate the effect of the number of beacons and their distance from the central FSO station. The beacons are positioned to uniformly surround the FSO station.
We consider the cases of 8 and 16 beacons, with radii 250 m, 500 m, 750 m, and 1 km. Each radius of the beacon circle requires a different range of $\tau$ values, as shown in Table \ref{tab:numeroffeat}. These values are deduced from the range of the satellite heights in our simulation. The temporal resolution of our simulation is 1 sec.
\begin{table}[h]
\footnotesize
\caption{Number of training features for each scenario.}
    \centering
    \begin{tabular}{|c|c|c|c|} 
    \hline
       Radius  &  $\tau$ values   (sec) & Beacons& Total features\\ 
            \hline
 1000 m&  $\tau_{1\text{--}6}=[28, 30, 32, 34, 36, 38 ]$&  16& 101\\ 
 & & 8&  53 \\ 
    \hline
750 m&  $\tau_{1\text{--}6}=[21, 22, 24, 26,  27, 28 ]$& 16 & 101 \\ 
    &  &  8   &  53 \\ 
    \hline
500 m&  $\tau_{1\text{--}6}=[14, 15, 16, 17, 18, 19 ]$&   16 & 101\\ 
 & & 8 & 53 \\ 
    \hline
   250 m&  $\tau_{1,2,3,4}=[7, 8, 9, 10 ]$ &     16 & 69\\ 
   &  &   8 & 37\\ 
    \hline
    \end{tabular}
    \label{tab:numeroffeat}
\end{table}
The neural network training features are the cloud-induced attenuation values, assuming that the transmission level is known and fixed as a function of satellite position.
A feature is assigned for each beacon-$\tau$ pair. So, for example, the first scenario in Table \ref{tab:numeroffeat}, with 16 beacons and six $\tau$ values, has ($16\times6$) attenuation features, plus 3 position and 2 direction features, making a total of 101 training features. The rest of the values in the table reflect the same feature assignment strategy. The scenarios are simulated and evaluated, each with its neural network features.

The attenuation measurements are taken every 1 second for the satellites with elevation angle above $30^{\circ}$ and stored in the experience buffer sequentially. The experience buffer size is 5000, corresponding to an average of 83 minutes. The batch size for learning is 500, and the learning rate is $ 10^{-2}$. We used a basic neural network that is five hidden layers deep with 350  rectified linear unit (ReLU) activation functions each. 
 We should also mention that choosing a $\tau$ span bigger than the needed values gives less accurate training results. This is due to the fact that these observations are not directly related to the predictions and can be considered as noise. Hence,  we limit the $\tau$ span in the case of a 250 m radius to the four inputs $\tau_{1,2,3,4}=[7, 8, 9, 10]$.









\subsubsection{Compromises and tradeoffs}

The duration of lookahead prediction is inherently constrained and cannot be indefinitely extended, primarily due to the limitations imposed by prediction accuracy. This section delves into the rationale behind this limitation and provides numerical insights into the trends and interdependencies between prediction accuracy and lookahead duration. Additionally, we consider the number of RF beacons as a variable influencing this delicate balance.


Fig. \ref{fig:pred250} illustrates the FSO attenuation levels alongside their respective predictive values ahead of time for a simulated scenario incorporating 16 RF beacons arranged in a 250 m radius circle. This figure compiles data across multiple satellites, with the lower segment depicting each satellite's elevation angle and distance upon reaching the $30^o$ elevation horizon. Consequently, each sequence that begins at an elevation angle of $30^o$ ascends and then descends corresponds to an individual satellite. It's important to note that these values are chronologically separated, as multiple satellites are typically above the horizon simultaneously. To maintain clarity, the values are organized non-chronologically, grouping those from the same satellite and sequencing them based on their entry into the horizon.

For evaluation, we define the prediction accuracy parameter as the ratio of correct predictions that the attenuation passes a specific threshold to the total predictions.    
The prediction accuracy $A_{Pred}$ is determined with relation to threshold $L_{th}$, applied in the following formula:
\begin{equation}
A_{Pred} = \dfrac{1}{N} \sum_{n=t_s}^{t_s + N} \mathbf{I} \{ (L_{FSO} > L_{th})~ \Leftrightarrow ~ (Pred_{FSO} > L_{th})  \},
\end{equation}
where N represents the total number of observational samples, $\mathbf{I} \{\}$ is the indicator function, which equals 1 if the condition inside is true and 0 otherwise. The biconditional logical operator $\Leftrightarrow$ ensures that the value inside the indication function is true if the conditions $(L_{FSO} > L_{th})$, and $(Pred_{FSO} > L_{th})$ are both true or both false. This way, a sample is deemed accurate if both the actual attenuation $L_{FSO}$ and its prediction $Pred_{FSO}$ are concurrently above or below the threshold.

\begin{table}[htbp]
 \caption{ Prediction Accuracies for different scenarios.  }
 \begin{center}
\begin{tabular}{|c||c|c|c|c|}
\hline
           &  250 m   &  500 m   & 750 m     &  1000 m       \\ \hline
8 Beacons  & 75 \%       &    67 \%       &     62 \%       &     60 \%     \\
16 Beacons  & 86 \%      & 81 \%      &    73 \%     &  70 \%  \\ \hline
\end{tabular}
  \label{tab:predacc}
  \end{center}
\end{table}

We simulate the configurations with 16 and 8 beacons, spanning radii from 250 to 1000 m, with $L_{th}=30 dB$, N = 5000 samples, and $t_{s}=30000$. The results of the prediction accuracy values are listed in Table \ref{tab:predacc}. Observations indicate that configurations with a greater beacon count yield higher training outcomes. This is because more beacons can introduce higher resolution in the predicted attenuation as a function of satellite direction.
Furthermore, smaller circle radii, associated with smaller $\tau$ values, contribute to enhanced training and more accurate predictions. Using smaller circle radii can help reduce inaccuracies when dealing with variably changing cloud shapes, particularly when the cloud layer is moving in a direction that differs from the satellite movement or when its speed is comparable to the speed of the attenuation points on the cloud (the points of satellite-ground link intersection with the cloud layer).
\subsection{Future Work}
We should also mention that the proposed learning model enforces the same $\tau$ range used for all the satellites in a specific scenario. This is not optimal because a satellite with a higher altitude would benefit from the prediction earlier with a high $\tau$ value. But, because the model is centralized, a satellite has to wait for the whole $\tau$ range to be evaluated with the learning model. Meaning that the prediction for the whole constellation is limited by the lowest $\tau$ value in the range. 
Future work can present a distributed multi-agent learning model where each satellite has its own learning model, which is exactly tailored to its needs. This way each learning neural network can have a single $\tau$ value per satellite. 

Moreover, our proposed method has the potential to predict the availability of alternate routes through different satellites. This capability can encourage the development of advanced routing strategies that can plan routes in advance, based on the shape and movement of clouds, as well as the position of the satellites. 
 There is ongoing work between NASA and commercial service providers to develop the next generation of satellite communication.   These developments consist of both optical and RF communication systems \cite{NASA2022CollaborateSpaceComms}. Since accurate prediction of FSO attenuation is key to achieving low-latency paths from the satellite constellations to ground stations, LEO optical communication constellations such as Starlink \cite{9393372} may utilize our model for enhanced route-precalculation, automated downlink route discovery, and low latency.

\section{Conclusion}


FSO links are inherently vulnerable to atmospheric disruptions like clouds, fog, and haze. This paper introduces a hybrid RF/FSO model that utilizes RF signal power to predict FSO signal attenuation and subsequent link availability, forecasting up to $\tau$ seconds ahead. The model hinges on RF beacons encircling the FSO station, with a centralized machine-learning model employing temporal correlation to map RF data to future FSO link attenuation.
Utilizing an experience buffer, our model assimilates measurements from the LEO satellite constellation at different intervals preceding the prediction, catering to varying satellite altitudes and corresponding prediction durations.  Simulations incorporating an emulated cloud model shed light on the RF signal profiles and their temporal correlation with FSO signals and explored the impact of different RF beacon configurations.
Our investigation also highlighted a trade-off between prediction duration and accuracy, influenced by the proximity of surrounding beacons. Notably, the system reached optimal predictive accuracy with 16 RF beacons, outperforming configurations with fewer beacons.

Our research demonstrates the potential of a data-driven, machine learning-based approach in enhancing the reliability of FSO links, particularly in challenging atmospheric environments.
The fact that RF beacons require only low bandwidth means they can be implemented and deployed at relatively low cost. Furthermore, this approach is inherently scalable; if additional predictive accuracy is desired, it can be easily achieved by adding more RF beacons.

\ifCLASSOPTIONcaptionsoff
  \newpage
\fi

\bibliographystyle{IEEEtran}
\bibliography{Bibliography}

\end{document}